\documentclass[twocolumn, tighten, times, twocolappendix]{aastex7}

\usepackage{amsmath}

\begin{document}

\title{Deep Adaptive Optics Imaging Rules Out a Helium Star Companion to PSR J1928+1815}

\author[orcid=0000-0002-1386-0603,gname=Pranav,sname=Nagarajan]{Pranav Nagarajan}
\affiliation{Department of Astronomy, California Institute of Technology, 1200 E. California Blvd., Pasadena, CA 91125, USA}
\email[show]{pnagaraj@caltech.edu}  

\author[orcid=0000-0002-6871-1752,gname=Kareem,sname=El-Badry]{Kareem El-Badry}
\affiliation{Department of Astronomy, California Institute of Technology, 1200 E. California Blvd., Pasadena, CA 91125, USA}
\email{kelbadry@caltech.edu}  

\author[orcid=0000-0002-4544-0750,gname=Jim,sname=Fuller]{Jim Fuller}
\affiliation{Department of Astronomy, California Institute of Technology, 1200 E. California Blvd., Pasadena, CA 91125, USA}
\email{jfuller@caltech.edu} 

\author[orcid=0009-0008-2437-0184,gname=Yunlang,sname=Guo]{Yunlang Guo}
\affiliation{School of Astronomy and Space Science, Nanjing University, Nanjing 210023, People’s Republic of China}
\email{yunlang@nju.edu.cn} 

\author[orcid=0000-0002-3865-7265,gname=Thomas,sname=Tauris]{Thomas M. Tauris}
\affiliation{Department of Materials and Production, Aalborg University, Fibigerstr{\ae}de 16, 9220 Aalborg, Denmark}
\email{tauris@mp.aau.dk} 

\begin{abstract}

PSR J1928+1815 is a 10.55 ms millisecond pulsar in a 3.6 hr orbit with a massive ($1.0$--$1.6\,M_{\odot}$) companion that produces extended radio eclipses. The companion, proposed to be a stripped helium star, is undetected in optical and infrared surveys. We present deep near-infrared imaging using Keck/NIRC2 with laser guide star adaptive optics. No source is detected at the pulsar position down to a $5\sigma$ limit of $K_s \approx 21.3$. Using stripped-star atmosphere models and conservative extinction estimates, we show that any plausible helium star companion would have been detected, ruling out this interpretation. A massive white dwarf (WD) companion remains consistent with the non-detection. We consider two possible origins for the eclipses: (1) absorption in a wind driven by a young, hot WD, and (2) material ablated from the WD by the pulsar. The former can naturally arise following Case BB mass transfer, which produces $\sim 1.2\,M_\odot$ WDs capable of sustaining winds of $\dot{M} \gtrsim 10^{-12}$--$10^{-13}\,M_\odot\,{\rm yr}^{-1}$ for $\sim 10^4$--$10^5$ yr, sufficient to obscure the pulsar at GHz frequencies. The latter requires efficient coupling of the pulsar's spin-down luminosity to the companion to drive the needed mass loss, which may be difficult to achieve. If the eclipse is powered by a WD wind, the system is likely observed in a short-lived phase; alternatively, if the companion is an older WD, the origin of the eclipsing material remains unclear. The apparent uniqueness of PSR J1928+1815 is consistent with a short detectability lifetime, though formation rate estimates remain uncertain.

\end{abstract}



\section{Introduction}
\label{sec:intro}

Millisecond pulsars (MSPs) are rapidly rotating ($P_{\text{spin}} \lesssim 30$ ms) neutron stars (NSs) that were spun up by accretion from a binary companion \citep[e.g.,][]{bhattacharya_formation_1991, 2023pbse.book.....T}. Recently, \citet{yang_helium_2025} discovered PSR J1928+1815, a 10.55 ms MSP in a 3.6 hour orbit. The system's radio timing solution implies a companion of mass $1.0$--$1.6\,M_{\odot}$ for plausible MSP masses and orbital inclinations. Unlike most other similar systems, PSR J1928+1815 features a broad radio eclipse covering 17\% of the orbit. However, the companion is significantly more massive than the wind-ablated $\lesssim 0.4\,M_{\odot}$ companions in eclipsing ``spider'' binaries \citep{chen_tauris_2013}. Since a main sequence star of the required mass would not fit in a 3.6 hr orbit, \citet{yang_helium_2025} propose that the massive companion is a helium (He) star, with winds from the He star causing the observed radio eclipse. The companion was not detected at optical or infrared wavelengths in survey data, with limiting magnitudes ranging from $23.3$ in the Pan-STARRS $g$-band \citep{PanSTARRS} to $18.8$ in the UKIDSS $K$-band \citep{ukidss_2007}, but \citet{yang_helium_2025} report these limits to be consistent with a He star companion.

If the companion is indeed a He star, it would make PSR J1928+1815 the only known pulsar with a He star companion and a valuable example of a massive binary that has rather unambiguously undergone common envelope evolution. In the formation scenario proposed by \citet{yang_helium_2025}, the $\approx 6\,M_{\odot}$ red giant progenitor of the He star expanded sufficiently to engulf its NS companion, leading to an episode of unstable mass transfer and formation of a common envelope. Once the red giant's envelope was successfully ejected, what remained was a NS orbiting a stripped He star (i.e., the core of the red giant) in a close orbit. \citet{yang_helium_2025} propose that highly super-Eddington mass transfer (i.e., NS accretion rate $\gtrsim 10^4\times$ the Eddington limit) during common envelope evolution recycled the NS, weakening its magnetic field and spinning it up to the observed spin period of $10.55$ ms.

While intermediate-mass (i.e., $1$--$8\,M_{\odot}$) stripped He stars in binaries have been observed in the Magellanic Clouds \citep{drout_gotberg_2023, gotberg_properties_2023, ludwig_sums_2026, blomberg_intermediate_2026}, the population of such stripped stars is still small, with none confirmed in the Milky Way. Furthermore, there are as yet no intermediate-mass stripped stars known to have NS companions, or to have formed through common envelope evolution. Unfortunately, detection of the PSR J1928+1815 system in the optical is precluded by the large amount of dust extinction along the sight-line to the system ($A_V \approx 8.5$ mag at $d \approx 8$ kpc based on the 3D dust map of \citealt{green_2019}). In this work, we present deep, adaptive optics (AO)-assisted near-infrared follow-up imaging of PSR J1928+1815, achieving unprecedented depth and spatial resolution with the ultimate goal of determining the true nature of the unseen companion.

Recently, \citet{gong_alternative_2025} presented near-infrared observations of PSR J1928+1815 which did not detect the companion and yielded deeper limits than archival survey data. Their observations, which were obtained without AO, disfavored but did not rule out a stripped star in the system. They propose that the binary hosts a massive white dwarf (WD) that is ablated by the wind of its MSP companion, producing a ``haze'' around the WD that gives rise to the observed radio eclipses. We compare our observational limits and discuss their proposed scenario in Section \ref{sec:ablation}.

The remainder of this work is organized as follows. In Section~\ref{sec:data}, we describe the near-infrared images taken with NIRC2 on Keck and aided by the laser guide star AO system. In Section~\ref{sec:results}, we use our non-detection of a He star to place constraints on the properties of the unseen companion. In Section~\ref{sec:discussion}, we discuss implications for the nature of the secondary, the eclipse mechanism, and the formation history of this MSP binary. Finally, in Section~\ref{sec:conclusion}, we summarize our conclusions and provide directions for future follow-up.

\section{Data}
\label{sec:data}

\begin{figure*}
    \centering
    \includegraphics[width=\textwidth]{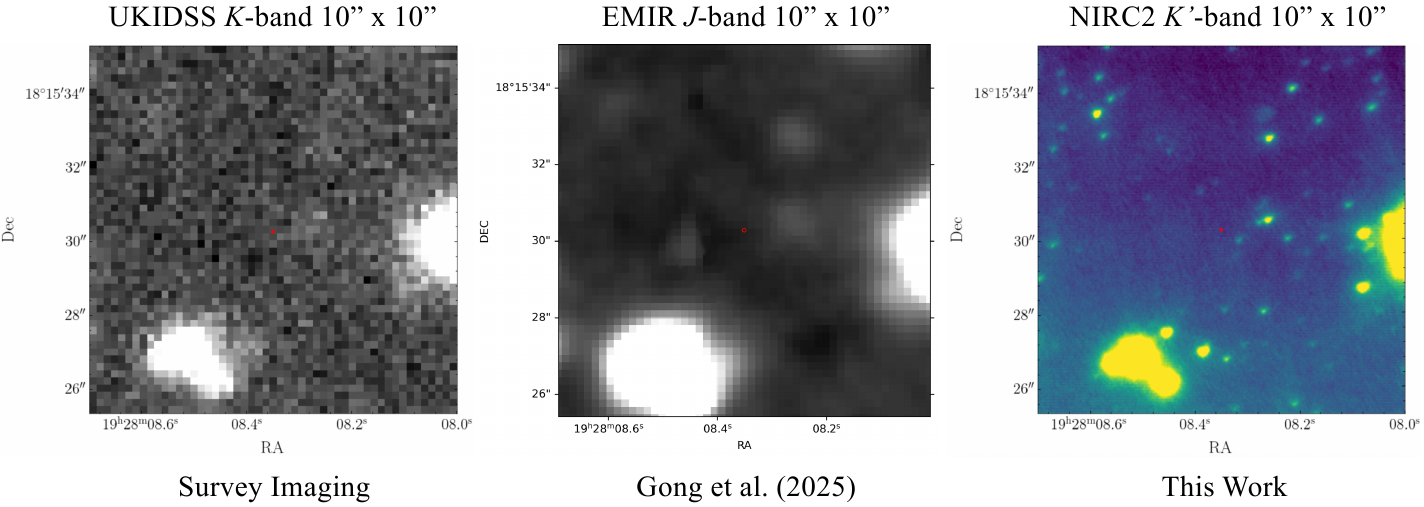}
    \caption{Comparison of 10'' $\times$ 10'' near-infrared cutouts centered on the radio localization of PSR J1928+1815 from the UKIDSS survey (left), the EMIR program of \citet{gong_alternative_2025}, and our NIRC2 campaign (right). In both the UKIDSS cutout and our image, the position of the pulsar is marked with a red circle, with the radius indicating the 0.02'' uncertainty in our WCS astrometric solution. In the EMIR image, the position of the pulsar is also marked with a red circle, but with a larger radius of 0.06'' corresponding to the mean astrometric accuracy reported by \citet{gong_alternative_2025}. The NIRC2 image is deeper and of higher spatial resolution than the UKIDSS and EMIR images, leading to the detection of many new sources. There is no near-infrared source detected at the position of PSR J1928+1815.}
    \label{fig:field_comp}
\end{figure*}

NIRC2 (PI: K. Matthews) is a near-infrared imager on the Keck II telescope designed to work with Keck adaptive optics (AO) to achieve high spatial resolution. We used the Keck laser guide star (LGS) AO system \citep{wizinowich_lgs_2006} for our observing campaign. Unlike with natural guide stars, the Keck LGS AO system is capable of working with tip-tilt stars as faint as $V = 18$ mag, opening up a much wider region of the sky to AO-assisted observations.

To maximize the field-of-view and to facilitate photometric calibration with a larger set of reference stars, we used the 40'' x 40'' wide camera, which has a pixel scale of 0.039686 arcsec/pixel. We used the K' filter, which has reduced background surface brightness relative to the standard K filter, enabling deeper imaging in the same integration time \citep[][]{wainscoat_cowie_1992}. The K' filter is also very similar to the Ks filter, allowing comparison to published 2MASS Ks-band magnitudes. Using three dither positions, we obtained 31 individual 30-second exposures of the field of PSR J1928+1815 over a period of $\approx 3.5$ hours on the night of June 28th, 2025 (UTC). Interruptions due to technical issues and satellite closures precluded full orbital phase coverage and caused the PSF quality to vary somewhat between sets of exposures. We reduced and stacked the individual images using the KAI pipeline\footnote{\texttt{github.com/Keck-DataReductionPipelines/KAI/tree/dev}} \citep{lu_kai_2021}, which performs dark subtraction, flat-fielding, hot pixel and cosmic ray removal, and image alignment. The code also accounts for the NIRC2 distortion solution \citep{service_lu_2016}. 

We show a 10'' x 10'' cutout of the stacked NIRC2 image of the field of PSR J1928+1815 in the right panel of Figure~\ref{fig:field_comp}, with the radio localization of the pulsar (RA = 19h 28m 08.349s, Dec = +18d 15m 30.27s) marked with a red circle. We compare our image against an equivalent 10'' x 10'' $K$-band cutout from the UKIRT Infared Deep Sky Survey (UKIDSS; \citealt{ukidss_2007}) in the left panel of Figure~\ref{fig:field_comp}. 

The UKIDSS Galactic Plane Survey has a depth of $K \approx 18.8$ \citep{ukidss_2007}.\footnote{All quoted magnitudes are in the Vega system.} Clearly, the NIRC2 data reaches a deeper magnitude limit and detects many new sources in the field unknown to prior survey imaging. Our imaging also resolves the sources detected in the UKIDSS image into several individual sources. However, we do not detect any source at the radio localization of PSR J1928+1815. The nearest candidate is a $K_s \approx 20$ point source located $\approx 0.5''$ to the southwest, well outside the total error circle of radius 0.02'' (see Section~\ref{sec:calibration}). Since this source is $\approx 25\sigma$ away from the location of the pulsar, it cannot be the near-infrared counterpart.


\section{Results}
\label{sec:results}

\subsection{Astrometric and photometric calibration}
\label{sec:calibration}

\begin{figure*}
    \centering
    \includegraphics[width=\textwidth]{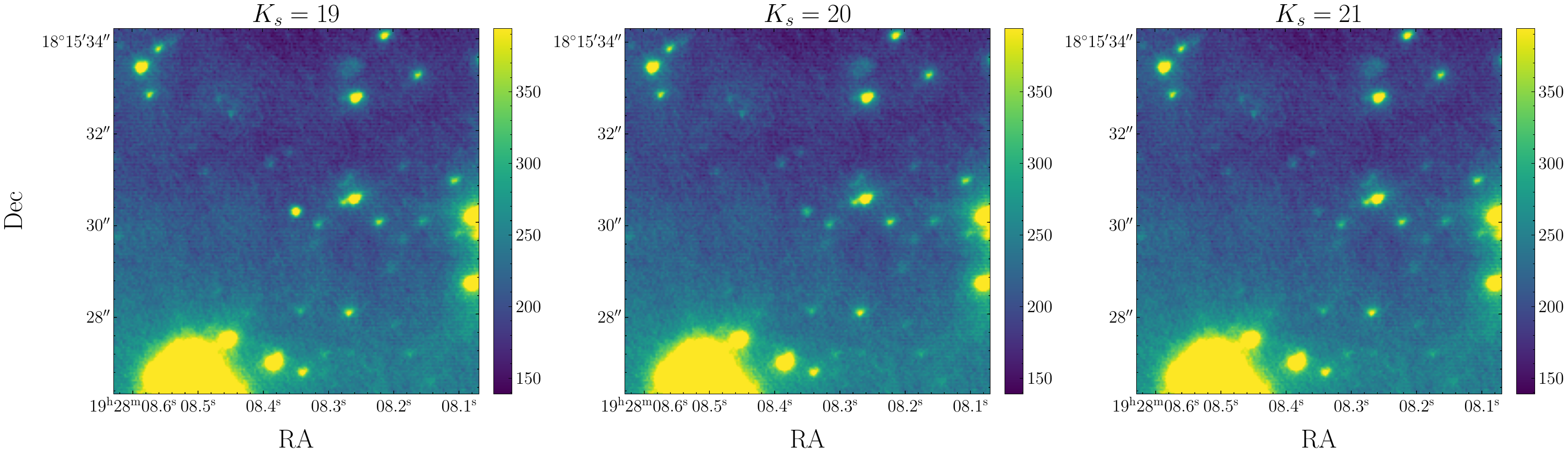}
    \caption{Injection and recovery tests of fake point sources of various $K_s$-band apparent magnitudes at the position of PSR J1928+1815 (located at the center of the cutout shown). Using aperture photometry, we determine the 5$\sigma$ limiting magnitude at the pulsar's location to be $K_s \approx 21.3$. The three panels show that a source of magnitude $K_s = 19$, $20$, or $21$ would be clearly detectable by eye at the position of the pulsar.}
    \label{fig:inject}
\end{figure*}

We use the published \textit{Gaia} DR3 right ascensions and declinations of nine bright reference stars in the 2MASS \citep{skrutskie_2006} catalog to derive a WCS astrometric solution. We estimated the uncertainty in the astrometric solution as the average discrepancy between the predicted and actual locations of the reference stars, yielding an uncertainty of 0.02 arcsec. The radius of the red circle in Figure~\ref{fig:field_comp} is not the error on the radio localization (0.01''; \citealt{yang_helium_2025}), but rather the adopted error in our astrometric solution, estimated to be 0.02''.

Next, we perform aperture photometry on reference stars with a range of brightnesses from the UKIDSS survey \citep{ukidss_2007} to derive a photometric calibration. Specifically, for each reference star, we sum all counts within an aperture radius of 5 pixels (roughly equivalent to the full-width half-max of non-saturated point sources in the image) centered on the star's location. Then, we subtract the local sky background, estimated as the sigma-clipped average within a concentric annulus of inner radius 8 pixels and outer radius 14 pixels. From these source fluxes and the known UKIDSS magnitudes, we derive an average zero-point to calibrate the empirical relation between instrument counts and apparent magnitude in the $K_s$ filter.

With this relation in hand, we perform an injection-and-recovery test to estimate the limiting magnitude at the location of the pulsar binary in our stacked image. We show injected sources of various apparent $K_s$-band magnitudes in Figure~\ref{fig:inject}. We model the point spread functions of the injected point sources as 2D circular Gaussians with a standard deviation of 2.0 pixels. We perform aperture photometry with the same aperture and annulus radii as before, defining the signal-to-noise ratio of a detection as the ratio between the source flux and the quadrature sum of source (Poisson) noise and sky noise. Based on this procedure, we find a $5\sigma$ limiting magnitude of $K_s \approx 21.3$. Figure~\ref{fig:inject} confirms that this result is reasonable; at $K_s = 21$, the injected source is faint, but still visually obvious.

\subsection{Constraints on He star companions}

Using the stripped star spectral models of \citet{gotberg_stripped_2018}, we predict the apparent $K_s$-band magnitude of $1.0$, $1.3$, and $1.6\,M_{\odot}$ He stars as a function of distance. For distances $\leq 8.0$ kpc, we apply extinctions from the 3D dust map of \citet{green_2019}. For distances $> 8.0$ kpc, we conservatively calculate the extinction as a linear interpolation between the value given by \citet{green_2019} at 8.0 kpc ($E(B-V) \approx 2.735$) and the total Galactic extinction given by the 2D dust map of \citet{schlafly_finkbeiner_2011} ($E(B-V) \approx 4.380$), which we take to apply at 12 kpc. We assume that $A_{K_s} = 0.306\,E(B-V)$ based on the \citet{fitzpatrick_extinction_1999} extinction law.

\begin{figure}
    \centering
    \includegraphics[width=\columnwidth]{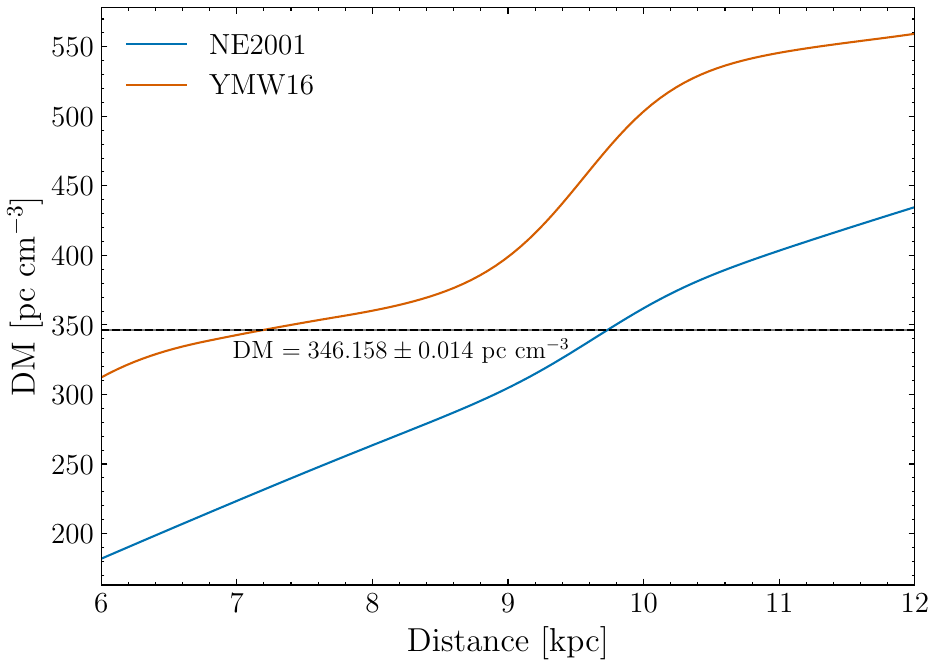}
    \caption{Predicted dispersion measure as a function of distance toward PSR J1928+1815 from the NE2001 \citep{cordes_model_2002} and YMW16 \citep{yao_electron_2017} Galactic electron density models. The dashed line and shaded band indicate the measured DM and its uncertainty. Intersections with the model curves imply distances of $7.2$ kpc (YMW16) and $9.7$ kpc (NE2001); we adopt a fiducial distance of $\approx8$ kpc.}
    \label{fig:dm}
\end{figure}

We show the constraints on the distance to PSR 1928+1815 from the electron density maps of \citet{cordes_model_2002} and \citet{yao_electron_2017} in Figure~\ref{fig:dm}. Specifically, we plot the predicted dispersion measure (DM) as a function of distance along the sight line to the pulsar, with the observed DM and corresponding uncertainty marked with a dashed black line and gray shaded region, respectively. Based on the observed DM of $346.158 \pm 0.014$ pc cm$^{-3}$ \citep{yang_helium_2025}, we would infer distances of $7.2$ kpc and $9.7$ kpc from the maps of \citet{yao_electron_2017} and \citet{cordes_model_2002}, respectively. We adopt a fiducial distance of $\approx 8$ kpc, which falls in-between these estimates.

\begin{figure*}
    \centering
    \includegraphics[width=\textwidth]{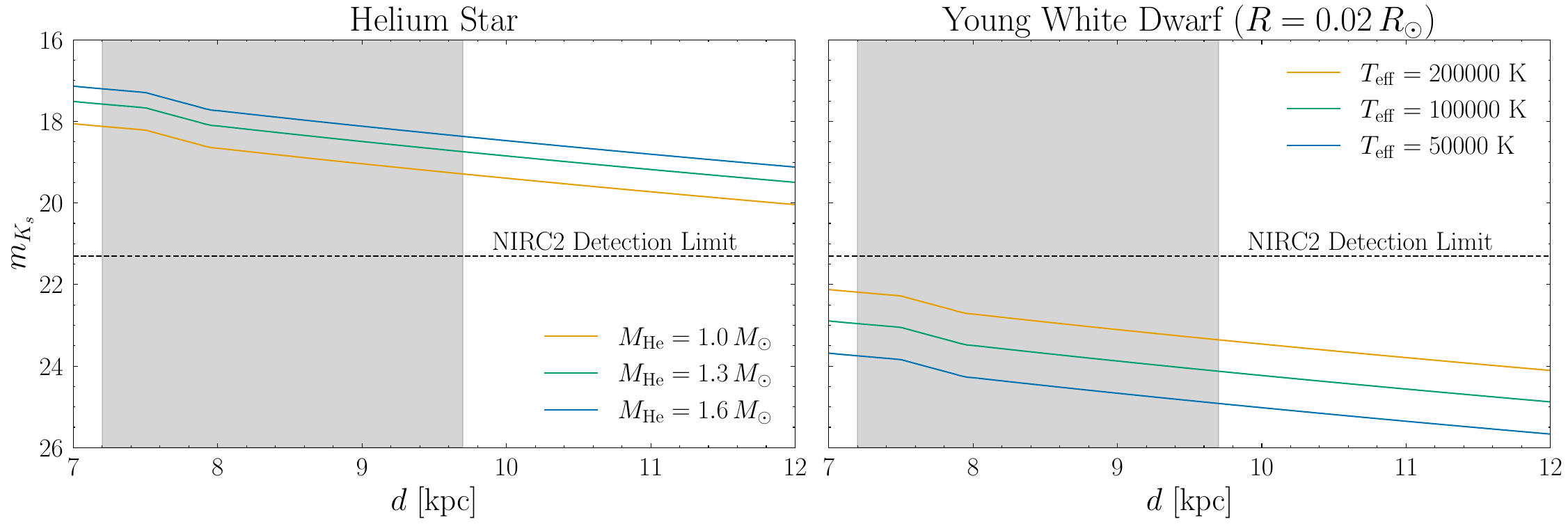}
    \caption{Predicted apparent $K_s$-band magnitudes of plausible stripped star (left) and white dwarf (right) companions as a function of distance. For the He stars, we adopt the stripped star spectral models of \citet{gotberg_stripped_2018}, while we assume blackbody radiation for the $0.02\,R_{\odot}$ WDs. For distances $\leq 8.0$ kpc, we use the extinctions from the 3D dust map of \citet{green_2019}, while for distances $> 8.0$ kpc, we linearly interpolate between the extinction at $8.0$ kpc and the total Galactic extinction from the 2D dust map of \citet{schlafly_finkbeiner_2011}, which we take to apply at 12 kpc. We shade the distance constraint based on the dispersion measure of \citet{yao_electron_2017} in gray. Based on our NIRC2 detection limit, we find that the would have detected any plausible He star, but would not have detected any plausible WD companion.}
    \label{fig:theoretical}
\end{figure*}

We display these predictions, along with our NIRC2 detection limit, in the left panel of  Figure \ref{fig:theoretical}. We shade the distance constraint from the pulsar's dispersion measure \citep{cordes_model_2002, yao_electron_2017} in gray. Even in the pessimistic case where all of the dust along the line of sight is in front of the He star (i.e., using the extinction from the dust map of \citealt{schlafly_finkbeiner_2011}), and the pulsar is at a distance of 12 kpc, we find that the He star should still have been detectable in our NIRC2-LGS image. Based on this non-detection, we rule out the He star hypothesis of \citet{yang_helium_2025}. 

Next, assuming blackbody radiation and a young white dwarf (WD) radius of $0.02\,R_{\odot}$\footnote{The typical radius of an old WD is $\lesssim 0.01\,R_{\odot}$. However, a young, hot (proto-)WD that has not yet settled onto the cooling track could be somewhat inflated.}, we predict the apparent $K_s$-band magnitudes of hot WDs of effective temperatures $2 \times 10^5$ K, $10^5$ K, or $5 \times 10^4$ K as a function of distance. We display these predictions, along with our NIRC2 detection limit, in the right panel of Figure \ref{fig:theoretical}. We find that, even in the case of a very hot and young WD, we would not have been able to detect a WD companion to PSR J1928+1815.

\subsection{Comparison to \citet{gong_alternative_2025}}

\citet{gong_alternative_2025} also pursued deep near-infrared imaging of PSR J1928+1815 using the EMIR instrument at the Gran Telescope Canarias (GTC). We show their EMIR $J$-band image in the middle panel of Figure~\ref{fig:field_comp}. Their follow-up image appears to be deeper than the UKIDSS $K$-band image, but has low spatial resolution relative to the NIRC2 $K'$-band image. In addition, many sources detected in the NIRC2 $K'$-band image are missing in the EMIR $J$-band image.

\citet{gong_alternative_2025} report $5\sigma$ detection limits in the $J$- and $H$-band of $23.7$ mag and $22.2$ mag, respectively. Assuming reasonable values for color (i.e., $J - K_s \approx 0$ and $H - K_s \approx 0$ for a He star) and extinction (i.e., values given in Table S2 of \citealt{yang_helium_2025} based on the 3D dust model of \citealt{marshall_2006}), our reported $K_s$-band limit of $21.3$ mag is comparable to these values. Specifically, the limits in the $J$-, $H$-, and $K_s$-bands are $\approx 3$ mag deeper than the predicted apparent magnitude of a $1.3\,M_{\odot}$ He star in each of these bands. However, our use of laser-guided AO corrections allows us to achieve much higher spatial resolution. In particular, we are able to resolve a nearby $K_s \approx 20$ source located about 0.5'' to the southwest of the target, which is not detected in the EMIR $J$-band image (see Figure~\ref{fig:field_comp}). There are many sources in the NIRC2 image that, assuming reasonable colors for infrared sources, are brighter than the reported limits of \citet{gong_alternative_2025}. The non-detection of these sources in the EMIR image suggests that the actual limits in that image may be less deep than \citet{gong_alternative_2025} report.

From their follow-up campaign, \citet{gong_alternative_2025} disfavor a He star companion. We agree with their hypothesis that the companion is instead likely to be a massive WD. However, we consider a scenario in which the companion is a young WD and the radio eclipse is wind-driven more favorable than the scenario they propose, in which the eclipse is caused by ablation of an old WD. We discuss these proposed scenarios further in Section~\ref{sec:discussion}. 

\section{Discussion} 
\label{sec:discussion}

\subsection{Nature of the companion}

Any hypothesis for the nature of the companion in PSR J1928+1815 has to explain the observed radio eclipse while also satisfying the derived constraints on the mass of the unseen secondary. Based on the mass range of $1.0$--$1.6\,M_{\odot}$ given by \citet{yang_helium_2025}, the companion to the MSP could be an (evolved) He star, another neutron star, or a massive white dwarf. We now consider each of these possibilities.

\subsubsection{He star}

\citet{yang_helium_2025} propose that the companion is a He star in the core-helium burning phase, stripped following an episode of common-envelope evolution. However, as we show in the left panel of Figure~\ref{fig:theoretical}, any plausible He star companion with mass between $1.0\,M_{\odot}$ and $1.6\,M_{\odot}$ would have easily been detected in our NIRC2 $K$-band observations of PSR J1928+1815. Hence, we discard this hypothesis. 

\subsubsection{Evolved He star}
\label{sec:evolved}

\citet{guo_eclipsing_2025} use the stellar evolution code Modules for Experiments in Stellar Astrophysics (\texttt{MESA}, version 10398; \citealt{paxton_2011, paxton_2013, paxton_2015, paxton_2018, paxton_2019, jermyn_2023}) to investigate the formation of eclipsing MSP binaries with He star companions. They explore a range of initial orbital periods ($0.04$--$2.00$ d) and companion masses ($0.5\,M_{\odot}$--$3.0\,M_{\odot}$), treating the NS as a $1.4\,M_{\odot}$ point mass and assuming a He star metallicity of $Z = 0.02$. They use the \texttt{co\_burn} nuclear reaction network along with Type 2 OPAL Rosseland mean opacity tables \citep{iglesias_rogers_1996}. They set the mixing-length parameter to 2.0 and the convective overshooting parameter to 0.014. They adopt the \citet{kolb_ritter_1990} mass transfer scheme and isotropic re-emission with $\alpha = 0$, $\beta = 0.5$, and $\delta = 0$ (i.e., they assume that half of the transferred mass is lost from the vicinity of the accretor; see e.g., \citealt{bhattacharya_formation_1991}). For more details, we refer the reader to \citet{guo_eclipsing_2025}.  

\begin{figure*}
    \centering
    \includegraphics[width=\textwidth]{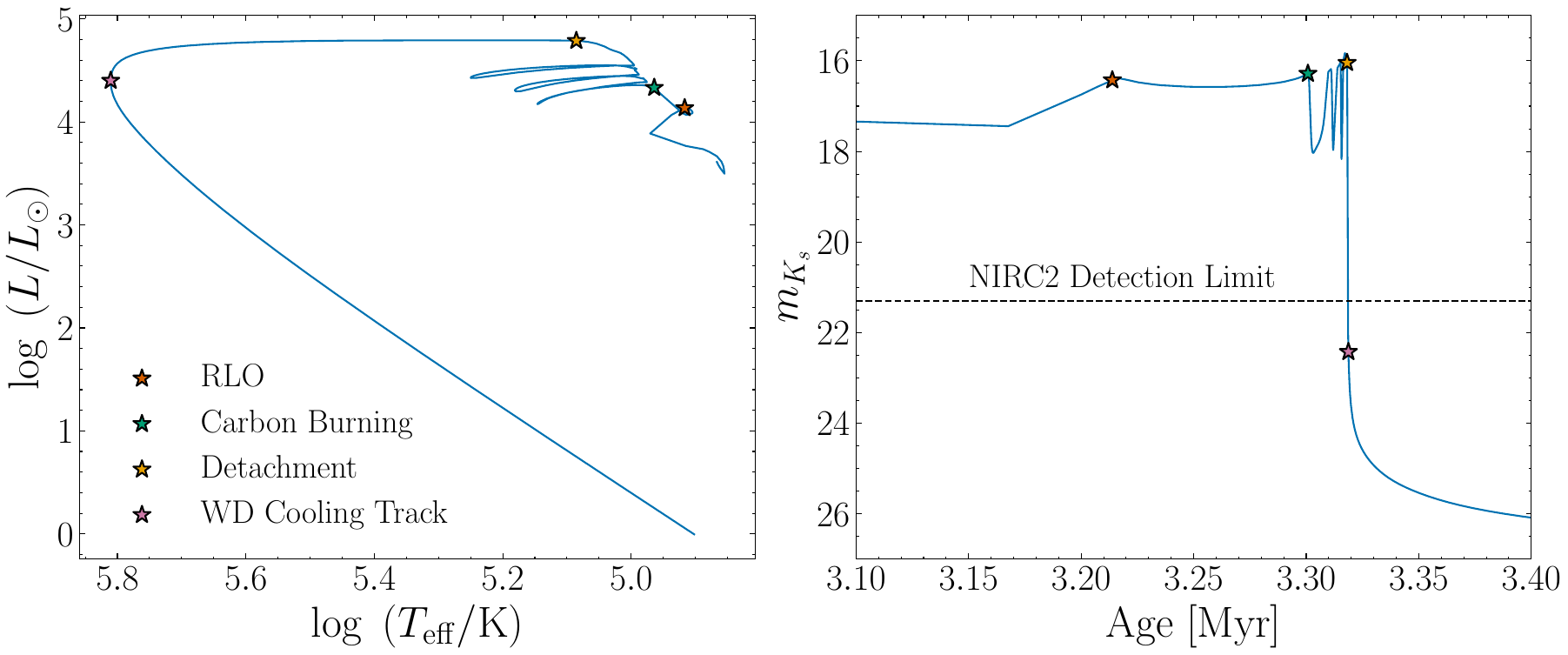}
    \caption{Left: Simulated evolution of a possible progenitor binary model for PSR J1928+1815 on a H-R diagram based on the fiducial \texttt{MESA} model of \citet{guo_eclipsing_2025}. The simulated binary has an initial orbital period of $\sim 0.1$ d and an initial He star mass of $2.2\,M_{\odot}$. The He star overflows its Roche lobe while undergoing helium shell burning, initiating a phase of Case BB mass transfer. The He star continues to evolve following the recycling of the NS, undergoing carbon shell flashes and brief periods of additional mass transfer before detaching completely and evolving down the WD cooling track. Right: Predicted $K_s$-band magnitude of the He star over time. Our NIRC2 detection limit is shown with a dashed line. Any He star would have been detected, during both He and C burning, with the $K_s$-band magnitude only following below the detection threshold when the companion turns onto the WD cooling track.}
    \label{fig:hr_mag}
\end{figure*}

\citet{guo_eclipsing_2025} perform simulations of both Case BA (i.e., Roche lobe overflow while the He star is still helium core burning) and Case BB (i.e., Roche lobe overflow while the He star is burning helium in a shell after core helium burning exhaustion) mass transfer scenarios for the formation of MSP + He star eclipsing binaries. From a dense grid of simulated binaries, they find that the binaries formed via Case BA mass transfer have short orbital periods of $0.01$--$0.05$ d, while those formed via Case BB mass transfer have longer orbital periods of $0.05$--$2.0$ d. They conclude that PSR J1928+1815 (current orbital period of $0.15$ d) likely formed via Case BB mass transfer in a binary with initial orbital period $\sim 0.1$ d and initial He star mass $\approx 2.2\,M_{\odot}$. They find that the Case BB Roche lobe overflow phase lasted for about $0.09$ Myr, sufficient to recycle the NS. Following this, the He star continues to evolve, undergoing carbon shell flashes before ending its life as a slowly cooling $\approx 1.2\,M_{\odot}$ ONe WD. We present the Hertzsprung-Russell (H-R) diagram evolution of the He star in the fiducial model of \citet{guo_eclipsing_2025} in the left panel of Figure~\ref{fig:hr_mag}.

\citet{guo_eclipsing_2025} suggest that, following detachment, irradiation from the MSP leads to evaporation of the He star, with the ablated material then causing the observed radio eclipse \citep[e.g.,][]{stevens_rees_podsiadlowski_1992}. In this phase, \citet{guo_eclipsing_2025} consider the companion to be an ``evolved'' He star, since the helium in its core has been exhausted. We consider the ablation mechanism further in Section~\ref{sec:ablation}. For now, we plot the evolution of the companion's predicted 2MASS $K_s$-band apparent magnitude over time in the right panel of Figure~\ref{fig:hr_mag}. In doing so, we model the He star as a blackbody (reasonable at near-infrared wavelengths), and assume a distance of $8$ kpc. We show our NIRC2 detection limit of $K_s = 21.3$ with a dashed line. We find that any evolved He star companion would have been detected, with the predicted $K_s$-band magnitude only falling below the detection threshold when the companion turns onto the WD cooling track. Indeed, the predicted $K_s$-band magnitude during the evolved He star phase is comparable to the magnitude during core helium burning. Thus, we can reject the evolved He star companion hypothesis.

\subsubsection{Neutron star companion}

If PSR J1928+1815 is a double neutron star binary consisting of a radio-quiet neutron star and a MSP, then the magnetosphere of the radio-quiet neutron star can eclipse the magnetosphere of the MSP, causing the observed dip at radio wavelengths (as has been observed in the double pulsar PSR J0737-3039, see e.g., \citealt[][]{breton_eclipse_2008}). However, the low eccentricity of the orbit ($e < 3 \times 10^{-5}$; \citealt{yang_helium_2025}) strongly suggests that the system did not receive a natal kick(s), making this scenario virtually impossible. In this scenario, even the loss in the companion's gravitational mass due to neutrino emission would result in a much larger eccentricity than that of the observed orbit. In detail, if the binary is initially circular, the post-supernova eccentricity due to symmetric mass loss (i.e., the Blauuw kick, \citealt{blaauw_1961}) is given by $e = \Delta M / M_{\text{final}}$, where $\Delta M$ is the instantaneous mass loss and $M_{\text{final}}$ is the total binary mass after the explosion \citep[e.g.,][]{hills_1983}. If the energy carried away by neutrinos during the supernova is $\Delta M c^2 \sim 10^{53}$ erg (i.e., comparable to the gravitational binding energy of the newly formed NS), the final eccentricity induced by this mass loss for a $1.4\,M_{\odot}$ + $1.4\,M_{\odot}$ double NS binary is $\approx 0.02$. This is a factor of $\sim 10^3$ larger than the observed orbital eccentricity of PSR J1928+1815, even before taking into account the contribution of the supernova ejecta to the total mass loss. Any asymmetry in the supernova would exacerbate the discrepancy, unless the resulting natal kick is extremely finely tuned \citep[e.g.,][]{brandt_effects_1995, vigna-gomez_constraints_2024}.

\subsubsection{White dwarf companion}

As shown in the right panel of Figure~\ref{fig:theoretical}, even a young, hot, massive WD with a radius of $\sim 0.02\,R_{\odot}$ and an effective temperature of $T_{\text{eff}} = 200,000$ K is predicted to be too faint to detect in our stacked NIRC2-LGS image. Indeed, a WD companion with mass between $1.0\,M_{\odot}$ and the Chandrasekhar limit satisfies the derived orbital constraints while also remaining consistent with our observational non-detection (see right panel of Figure~\ref{fig:hr_mag}). Based on our NIRC2 detection limit and the \texttt{MESA} simulations of \citet{guo_eclipsing_2025}, we consider a $\approx 1.2\,M_{\odot}$ ONe WD secondary to be the hypothesis that can most likely explain the data.

\subsection{Eclipse mechanism}
\label{sec:eclipse}

The remaining piece of the puzzle is the eclipse mechanism, as a typical massive WD has a predicted physical radius $\lesssim 0.01\,R_{\odot}$, which would only block the MSP's radio emission over $\approx 0.2\%$ of the orbital period in an edge-on orbit. Instead, the WD must have an extended gaseous atmosphere that causes the observed radio eclipse. Two possibilities for the source of this absorbing gas include ablation of the WD by irradiation from the MSP and/or a stellar wind from the WD itself. We now consider each of these mechanisms in turn.

\subsubsection{Ablation of the companion}
\label{sec:ablation}


\citet{gong_alternative_2025} suggest that the massive WD is ablated by the MSP, causing the radio eclipse observed today. In doing so, they relate the WD ablation rate to the pulsar's total spin-down luminosity\footnote{In Table 1 of their paper, \citet{gong_alternative_2025} report the spin-down age of the MSP as 0.46 Gyr. This is likely a typo, since the actual spin-down age is about 46 Myr, a factor of $10$ smaller \citep{yang_helium_2025}. However, this does not appear to affect any of their subsequent calculations. The error has since been corrected in the erratum by \citet{2026ApJ..1001..126G}.} of $1.2 \times 10^{35}$ erg s$^{-1}$ \citep{yang_helium_2025}, which they point out is much higher than for other typical MSPs. Instead, it is generally only the pulsar's $\gamma$-ray luminosity $L_{\gamma}$ that couples to the companion's upper atmosphere, driving an ablated wind via Compton heating \citep[e.g.,][]{ginzburg_quataert_2020}. In detail, the ablation rate $\dot{M}$ is given by:

\begin{equation}
    \frac{G M \dot{M}}{R} \equiv -\eta L_{\gamma} \left(\frac{R}{a}\right)^2,
\end{equation}

\noindent where $M$ is the companion's mass, $R$ is the companion's radius, $a$ is the orbital separation, and $\eta$ is the fraction of the incident energy that efficiently evaporates the companion \citep[e.g.,][]{ginzburg_quataert_2020}. \citet{gong_alternative_2025} adopt $\eta = 0.05$, which is likely too high. Instead, \citet{ginzburg_quataert_2020} show that the evaporation efficiency can be estimated as:

\begin{equation}
    \eta \sim 2.2 \times 10^{-4} \left(\frac{L_{\gamma}}{L_{\odot}}\right)^{1/3} \left(\frac{M}{10^{-2}\,M_{\odot}}\right)^{1/9} \left(\frac{P_{\text{orb}}}{1 \text{ h}}\right)^{-2/9}.
\end{equation}

As an upper limit, and to facilitate comparison to \citet{gong_alternative_2025}, we set the pulsar's $\gamma$-ray luminosity equal to its total spin-down luminosity. Then, using a WD mass of $M_{\text{WD}} = 1.2\,M_{\odot}$ and an orbital period of $P_{\text{orb}} = 0.15$ d, we find that $\eta \approx 8.9 \times 10^{-4}$. For a typical NS mass of $M_{\text{NS}} = 1.4\,M_{\odot}$, Kepler's Third Law implies an orbital separation of $a \approx 1.6\,R_{\odot}$. Adopting a typical WD radius of $R_{\text{WD}} = 0.01\,R_{\odot}$, we derive an upper limit on the ablation rate of $\dot{M} \approx 3 \times 10^{-16}\,M_{\odot}$ yr$^{-1}$. 

Even with these optimistic assumptions, the calculated upper limit is an order of magnitude lower than the ablation rate of $4.55 \times 10^{-15}\,M_{\odot}$ yr$^{-1}$ derived by \citet{gong_alternative_2025}. This, in turn, increases the formation timescale of the radio-opaque ``haze'' surrounding the WD by an order of magnitude, making it longer than the haze's cooling timescale \citep[e.g.,][]{gong_alternative_2025}. The haze cools faster than the ablation occurs, and radio eclipses would no longer be expected. These conditions challenge ablation of the WD as the most likely eclipse mechanism.

\subsubsection{Wind from a young WD}
\label{sec:WD_wind}

Next, we consider winds from a young, hot WD companion to PSR J1928+1815 as an alternative hypothesis. GHz radio waves can be attenuated by either free-free absorption or synchrotron absorption in the WD wind. We show in Appendix~\ref{sec:freefree} that free-free absorption is likely insufficient to explain the observed eclipse.

On the other hand, synchrotron absorption is a viable eclipse mechanism. In this scenario, the magnetic field arises from the motion of charged particles in the pulsar wind at the location of the intrabinary bow shock. The characteristic magnetic field is given by equating the plasma magnetic energy density $B^2 / (8 \pi)$ to the pulsar wind energy density $\dot{E}/(4 \pi c a^2)$, where $\dot{E}$ is the pulsar's spin-down luminosity and $a$ is the orbital separation. Using $\dot{E} = 1.2 \times 10^{35}$ erg s$^{-1}$ and $a \approx 1.6\,R_{\odot}$, we estimate $B \approx 25$ G. The cyclotron frequency is $eB / (2 \pi m_e c) = 70$ MHz. At an observing frequency of 1.25 GHz, the cyclotron harmonic $m$ is about 18, implying that absorption occurs in the synchrotron regime.

The optical depth of synchrotron absorption due to a population of non-thermal electrons with power law index $p$ is given by \citep{thompson_physical_1994}:

\begin{equation}
\begin{aligned}
\tau =
\left(\frac{3^{\frac{p+1}{2}} \Gamma\left(\frac{3p+2}{12}\right) \Gamma\left(\frac{3p+22}{12}\right)}{4}\right) \\ \left(\frac{\sin{\theta}}{m}\right)^{\frac{p + 2}{2}} \frac{n_0 e^2}{m_e c \nu} \mathcal{L},
\end{aligned}
\end{equation}

\noindent where $\theta$ is the angle between the $B$-field and the line of sight, $n_0$ is the number density of non-thermal electrons, and $\mathcal{L}$ is the path length through the absorbing medium. We assume that the WD wind consists of fully ionized helium. Then, for a relativistic electron fraction $f$ and a shock compression factor $\chi$, we can estimate $n_0$ at the apex of the bow shock to be:

\begin{equation}
    n_0 \sim \chi\,f\,\left(\frac{\dot{M}}{8 \pi R_0^2 v_w m_p}\right),
\end{equation}

\noindent where $\dot{M}$ is the WD wind mass loss rate, $v_w$ is the wind speed, and $R_0$ is the minimum distance from the WD to the bow shock. This minimum distance is given by \citep{canto_exact_1996}:

\begin{equation}
    R_0 = \frac{a \sqrt{\eta_w}}{1 + \sqrt{\eta_w}},
\end{equation}

\noindent where $\eta_w$ is the wind momentum ratio between the WD and the MSP:

\begin{equation}
    \eta_w \equiv \frac{\dot{M} v_w c}{\dot{E}}.
\end{equation}

For a mass loss rate of $\dot{M} \sim 10^{-12}\,M_{\odot}$ yr$^{-1}$ (i.e., the approximate wind limit, see Section \ref{sec:wind}) and a wind speed of $v_w \sim 10^4$ km s$^{-1}$ (i.e., the escape velocity of a $1.2\,M_{\odot}$ WD), we find a wind momentum ratio of $\eta_w \approx 0.016$ and a minimum distance from the WD to the bow shock of $R_0 \approx 0.18\,R_{\odot}$. 

In the strong-shock limit, $\chi = 4$. If we assume a typical relativistic fraction (at the low end) of $f \sim 0.01$ \citep{thompson_physical_1994}, then the non-thermal electron number density $n_0 \approx 3.9 \times 10^5$ cm$^{-3}$. For simplicity, we take the binary to be edge-on (i.e., $\sin i \approx 1$) and at conjunction, with the WD between the MSP and the observer, so that $\mathcal{L} \sim R_0$. Adopting a power law index $p = 2.5$ \citep{thompson_physical_1994}, we find that $\tau > 1$ for $\theta \gtrsim 6.6^{\circ}$, and that $\tau > 10$ for $\theta \gtrsim 18.7^{\circ}$. In other words, we find that $\tau \gg 1$ at $L$-band radio frequencies for plausible geometries and typical wind parameters. Furthermore, if we instead adopt a weaker wind mass loss rate of $\dot{M} \sim 10^{-13}\,M_{\odot}$ yr$^{-1}$, we still find that $\tau > 10$ for $\theta \gtrsim 33.7^{\circ}$. We conclude that synchrotron absorption of the pulsar's radio emission by a young, hot (proto-)WD's wind is likely sufficient to explain the observed radio eclipse.

The geometry of the bow shock is given (to first order) by the exact solution of \citet{canto_exact_1996}. The asymptotic opening angle $\theta_{\infty}$ of the bow shock (measured from the line connecting the WD and the MSP) is given by \citet{canto_exact_1996}:

\begin{eqnarray}
    \theta_{\infty} - \tan{\theta_{\infty}} = \frac{\pi}{1 - \eta_w}.
\end{eqnarray}

For our (somewhat pessimistic) estimate of $\eta_w \approx 0.016$, $\theta_{\infty} \approx 150.5^{\circ}$. If the orbit is edge-on and the shock axis of symmetry lies in the orbital plane, then the maximal eclipse fraction is $f_E = (\pi - \theta_{\infty})/\pi \approx 16.4$\%, in good agreement with the observed eclipse fraction of $\approx 17\%$. Realistically, the bow shock is not symmetric about conjunction, and the shock surface is not always optically thick \citep[e.g.,][]{wadiasingh_constraining_2017}. A full assessment therefore requires a self-consistent numerical calculation of the shocked-wind geometry and radio opacity, which we defer to future work. That being said, the optical depth and eclipse duration should increase with decreasing frequency. This provides a potential observational test of our hypothesis.


\subsection{Binary evolution modeling}

\subsubsection{Duration of the WD wind}
\label{sec:wind}

How long would a WD be able to drive winds strong enough to cause a radio eclipse? PG 1159 stars, which have lower $\log g$ values than the massive white dwarfs considered here, are thought to evolve into DO or DAO white dwarfs when they cross the ``wind limit'' at mass loss rates of $\dot{M} \sim 10^{-12}$--$10^{-13}\,M_{\odot}$ yr$^{-1}$ \citep[e.g.,][]{unglaub_hot_2000}. Observations of PG 1159 stars show strong evidence confirming the existence of this limit as well \citep[e.g.,][]{mackensen_overweight_2025}. For low-mass PG 1159 stars, the typical timescale to cross the wind limit is $\sim 10^6$ yr \citep[e.g.,][]{unglaub_hot_2000}. The wind limit is both observationally and theoretically uncertain at higher WD masses. To estimate how long it would take a (proto-)WD to cross the extrapolated wind limit, we estimate the wind mass loss rate by applying empirical wind prescriptions to the \texttt{MESA} models discussed in Section~\ref{sec:evolved}.

\citet{guo_eclipsing_2025} use \texttt{MESA} to simulate the evolution of a grid of PSR J1928+1815-like binaries past detachment, terminating the code when the He star companion evolves into an CO or ONe WD and cools down to a luminosity of $1\,L_{\odot}$. Adopting the stellar parameters of the companion from their model with initial orbital period $0.1$ d and initial He star mass $2.2\,M_{\odot}$, we use the \citet{jeffery_hamann_2010} prescription for extreme He stars to estimate the wind mass loss rate from the luminosity:

\begin{equation}
    \log \dot{M} = 1.5 \log{(L/L_{\odot})} - 14.4.
\end{equation}






\begin{figure*}
    \centering
    \includegraphics[width=0.9\textwidth]{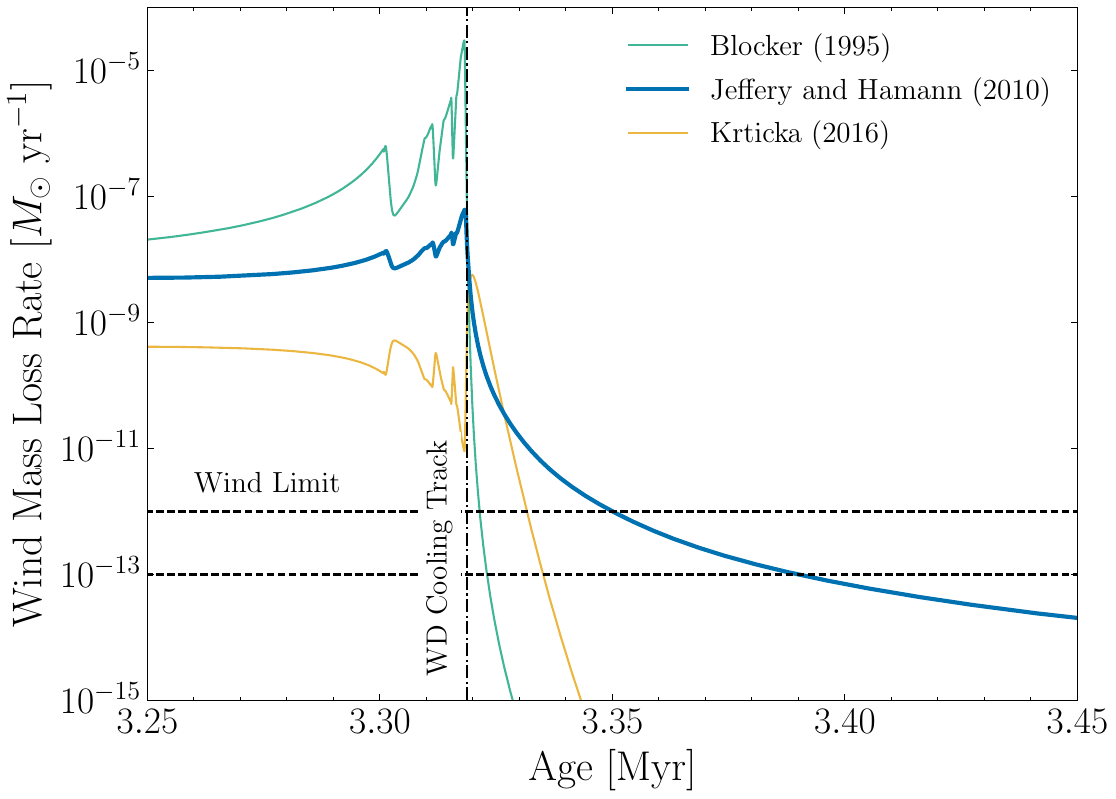}
    \caption{Predicted wind mass loss rate over time for the young (proto-)WD companion in the fiducial \texttt{MESA} model of \citet{guo_eclipsing_2025}, based on the prescription of \citet{jeffery_hamann_2010} for extreme He stars. We mark the wind limit, below which the WD wind is expected to be negligible, with a black dashed line. Once the proto-WD turns onto the WD cooling track (dot-dashed line), it is able to launch a wind $\gtrsim 10^{-12}$--$10^{-13}\,M_{\odot}$ yr$^{-1}$ for $10^4$--$10^5$ yr. However, this time scale is uncertain; to emphasize this, we use lines of lower opacity to show predicted mass loss rates based on alternative wind prescriptions for post-AGB stars \citep{bloecker_wind_1995} and hot subluminous stars \citep{krticka_wind_2016}. These prescriptions imply detectability timescales that are reduced by up to an order of magnitude; on the other hand, they are not as applicable as the prescription of \citet{jeffery_hamann_2010} for modeling the wind of a hot (proto-)WD.}
    \label{fig:wind}
\end{figure*}

We plot the predicted wind mass loss rate of the young WD over time in Figure~\ref{fig:wind}. We find that the proto-WD can launch the wind required to produce the observed radio eclipse for $\approx 10^4$--$10^5$ yr. After this point, the predicted mass loss rate drops below the wind limit of $\sim 10^{-12}$--$10^{-13}$ $M_{\odot}$ yr$^{-1}$, and the wind launched by the WD is expected to become negligible \citep[e.g.,][]{unglaub_hot_2000}. The eclipsing timescale we derive for this scenario is comparable to the $\sim10^4$ yr timescale suggested by \citet{yang_helium_2025} based on observations of helium WDs, though we emphasize the uncertainty in extrapolating wind prescriptions for stripped stars or spider binaries to the parameter space of interest. To visualize this uncertainty, we show predicted mass loss rates based on wind prescriptions for post-AGB stars \citep{bloecker_wind_1995} and hot subluminous stars \citep{krticka_wind_2016} in Figure~\ref{fig:wind}. Compared to the wind prescription of \citet{jeffery_hamann_2010}, these alternative prescriptions imply detectability timescales that are smaller by up to an order of magnitude; on the other hand, they are not as applicable for modeling the wind of a hot (proto-)WD.

\subsubsection{Recycling of the NS}

We now review the plausible mass transfer scenarios that can explain the formation of the present-day recycled pulsar-WD binary.

\paragraph{Pre-common envelope}

In this scenario, the donor was an intermediate-mass giant star separated from the NS by $\gtrsim 3$ AU. Initially, the NS accreted from the wind of the giant, and the system would have been observable as a symbiotic X-ray binary \citep[e.g.,][]{yungelson_2019}. While wind accretion is likely insufficient to recycle the NS to millisecond spin periods \citep[e.g.,][]{lu_population_2012}, the giant eventually filled its Roche lobe, initiating unstable mass transfer. \citet{deng_formation_2025} propose that PSR J1928+1815-like systems can be formed if super-Eddington accretion via Roche lobe overflow occurred prior to the common envelope phase. Following the ejection of the giant's envelope, what was left behind was a recycled pulsar orbiting a He star or a young, massive WD in a tight orbit. 

This scenario requires some degree of fine tuning. Only a small region of the parameter space of initial conditions that \citet{deng_formation_2025} explore allows for both successful ejection of the common envelope and recycling of the NS to spin periods $< 30$ ms while also avoiding a merger event.

\paragraph{During common envelope}

\citet{nie_formation_2026} study the formation of MSPs in low-mass or intermediate-mass X-ray binaries that undergo common-envelope evolution, performing an extensive grid of \texttt{MESA} simulations for a range of initial orbital periods, donor masses, and common envelope ejection efficiencies. They report that two distinct evolutionary channels with initial donor mass $\approx 6.0\,M_{\odot}$ can produce PSR J1928+1815-like systems. In the first scenario, $\alpha_{\text{CE}} = 3.0$ or $1.0$, and the binary undergoes Case B mass transfer prior to the common-envelope phase. In the second scenario, $\alpha_{\text{CE}} = 0.3$, and the binary begins in a wider orbit and undergoes unstable Case C mass transfer instead. In both cases, they propose that the pulsar was recycled by accreting $\sim 0.01\,M_{\odot}$ during the common-envelope evolution phase.\footnote{\citet{yang_helium_2025} also propose that the NS was recycled during the common-envelope evolution phase. This requires the NS accretion rate to be highly super-Eddington (i.e., $\gtrsim 10^4$ times the Eddington limit), which might be possible due to neutrino cooling \citep[e.g.,][]{houck_hypercritical_1991, macleod_accretion_2015}.}

\citet{nie_formation_2026} consider the second scenario to be consistent with the massive WD hypothesis. In that scenario, the binary shrinks to the observed orbital period via gravitational wave radiation following the common-envelope detachment phase, providing enough time for the He star to evolve into a WD. However, in their model, the post-common envelope binary requires $\approx 3.65$ Gyr for the orbit to shrink to the observed orbital period of $0.15$ d. The fact that this is significantly longer than the 46 Myr spin-down age of the MSP disfavors this hypothesis, though the final orbital period after common envelope evolution is uncertain.

\paragraph{Post-common envelope}

This is the scenario in the fiducial \texttt{MESA} model that we consider \citep{guo_eclipsing_2025}. In this case, the donor was a $\gtrsim 10\,M_{\odot}$ massive star separated from the NS by $\sim 1$ AU. The massive star filled its Roche lobe before burning through its core helium, leading to unstable Case B mass transfer. Following common-envelope evolution, the binary emerged as a $\approx 2.2\,M_{\odot}$ He star orbiting a NS in a tight orbit. After completing core helium burning, the He star expanded, eventually filling its Roche lobe and initiating stable Case BB mass transfer. This episode of mass transfer recycled the NS into a pulsar with spin period $\sim 10$ ms. Finally, the binary detached and the He star evolved into a young, massive WD, producing the binary we observe today.

Theoretical simulations show that Case BB mass transfer following common envelope evolution can readily form binaries featuring massive CO or ONe WDs orbiting MSPs in close orbits \citep[e.g.,][]{dewi_naked_2002, tauris_formation_2012, lazarus_recycled_2014, guo_eclipsing_2025}.\footnote{PSR J1952+2630, which features a massive WD in a 9.4 hr orbit around a 20.7 ms recycled pulsar \citep{lazarus_recycled_2014}, is similar to PSR J1928+1815. PSR J1952+2630 also has a small spin-down age \citep[77 Myr;][]{lazarus_recycled_2014}, but does not show radio eclipses.} In their fiducial model, \citet{guo_eclipsing_2025} find that the Case BB RLO lasted for $\approx 0.09$ Myr, with the mass transfer rate being highly super-Eddington (i.e., $\sim 10^3$ times the Eddington limit). Based on this model, we compute that the donor loses a total mass of $\approx 0.97\,M_{\odot}$ over the mass transfer history of PSR J1928+1815. Since just $\sim 0.01\,M_{\odot}$ of material is sufficient to spin up the NS to a $10.55$ ms spin period \citep[e.g.,][]{tauris_formation_2012}, the accreted mass fraction (i.e., spin-up efficiency) must be $\sim 10^{-2}$ to be consistent with observations. Indeed, a NS accretion rate that is a factor of $\sim 3\times$ the Eddington limit is sufficient to recycle the pulsar in this scenario \citep[e.g.,][]{tauris_2017, guo_eclipsing_2025}.



\subsection{Is the detectability lifetime consistent with the detection of PSR J1928+1815?}

From Monte Carlo binary population synthesis modeling, \citet{guo_eclipsing_2025} estimate the Galactic formation rate for eclipsing MSP + (evolved) He star binaries formed via Case BB mass transfer (which will eventually evolve into MSP + WD binaries) to be $R _{\text{form}} \approx 2 \times 10^{-4}$ yr$^{-1}$.\footnote{\citet{yang_helium_2025} derive a lower formation rate of ($1.3$--$7.2$) $\times 10^{-6}$ yr$^{-1}$. This is likely because they define their target population as having a narrower range of companion masses and orbital periods. They also assume a lower star formation rate of $3\,M_{\odot}$ yr$^{-1}$ for 10 Gyr, though we account for this uncertainty in our final estimate.} In doing so, they assume a \citet{miller_scalo_1979} initial mass function and a uniform mass ratio distribution. They also assume that all stars are in circularized binaries, with the distribution of orbital separations being uniform for wide binaries and falling off smoothly for close binaries \citep[e.g.,][]{han_bps_2020}. They combine the common envelope ejection efficiency parameter $\alpha_{\text{CE}}$ and structure parameter $\lambda$ into the free parameter $\alpha_{\text{CE}} \lambda$, which they set to either $0.5$ or $1.0$. Finally, they assume a constant Milky Way star formation rate of $5\,M_{\odot}$ yr$^{-1}$ over the past 15 Gyr. For more details, we direct the reader to \citet{guo_eclipsing_2025}. 

Based on our results from Section~\ref{sec:wind}, let us suppose that the detectability timescale of PSR J1928+1815 (i.e., the lifetime over which the young WD launches a wind with a mass loss rate substantial enough to cause a radio eclipse) is $t_{\text{det}} \sim 5 \times 10^4$ yr. To estimate the number of expected detections of PSR J1928+1815-like systems, we can write:

\begin{eqnarray}
    N = R _{\text{form}} \times t_{\text{det}} \times f_{\text{beam}} \times f_{\text{eclipsing}} \times f_{\text{complete}},
\end{eqnarray}

where $f_{\text{beam}}$ is the beaming fraction, $f_{\text{eclipsing}}$ is the fraction of binaries with inclinations such that a radio eclipse is observable, and $f_{\text{complete}}$ accounts for survey completeness. The beaming fraction of MSPs is believed to be in the range of $0.4 < f_{\text{beam}} < 1$, and likely falls close to unity \citep[e.g.,][]{levin_msp_2013}. Geometrically, for an eclipse to be observed, the inclination must be greater than $i_{\min} = \sin^{-1}\left(-\cos{\theta_{\infty}}\right) \approx 60.5^{\circ}$ \citep[e.g.,][]{wadiasingh_constraining_2017}. The eclipse probability is then $f_{\text{eclipsing}} = \cos{i_{\min}} \approx 0.5$. Finally, MSPs have a scale height of 500 pc \citep[e.g.,][]{levin_msp_2013}, so the vast majority of them are found within $10^{\circ}$ of the Galactic Plane, within the footprint of the Galactic Plane Pulsar Snapshot (GPPS) survey. As of mid-2025, the GPPS survey was $\approx25$\% complete in terms of sky area coverage \citep[][]{fast_four_2025}. However, the survey has already covered the majority of the planned sky area within a few degrees of the Galactic Plane, where most MSPs, and particularly young systems like this PSR J1928+1815, are found \citep[][]{fast_four_2025}. Furthermore, since PSR J1928+1815 is $\sim 8$ kpc away, the GPPS survey is likely complete to most of the Milky Way within the region it has observed. Adopting $f_{\text{beam}} \sim 1$, $f_{\text{eclipsing}} \sim 0.5$, and $f_{\text{complete}} \sim 0.5$, we find that $N \approx 2.5$.

An important caveat is that the estimated formation rate of (eclipsing) MSP + (evolved) He star binaries from population synthesis modeling is uncertain. For instance, the star formation rate assumed by \citet{guo_eclipsing_2025} is likely too high; the current star formation rate of the Milky Way is closer to $\sim 2\,M_{\odot}$ yr$^{-1}$ \citep[e.g.,][]{licquia_newman_2015}. Nevertheless, even after accounting for this overestimate, $N \sim 1$, consistent with the unique detection of PSR J1928+1815.

It may be objected that the characteristic spin-down time of the pulsar, which is 46 Myr, is much longer than our adopted detectability timescale of $\sim 10^4$--$10^5$ kyr. However, our proposed formation scenario {\it requires} the MSP to be young, since otherwise its WD companion would not be able to launch a $\gtrsim 10^{-12}$--$10^{-13}\,M_{\odot}$ yr$^{-1}$ wind, and the system would not be discovered as eclipsing. That is, if radio eclipses are due to a wind from a hot WD, eclipsing MSP + WD systems will always have true ages much smaller than their spin-down timescales. While our proposed scenario requires observing PSR J1928+1815 in a short-lived phase of binary evolution, the origin of the eclipse remains difficult to explain for an older system (see Section~\ref{sec:ablation}).


\section{Conclusion}
\label{sec:conclusion}

PSR J1928+1815 is a 10.55 ms millisecond pulsar (MSP) in a 3.6-hr orbit discovered via radio timing by \citet{yang_helium_2025}. Since the companion mass is $1.0$--$1.6\,M_{\odot}$ and the pulsar is eclipsed in the radio, \citet{yang_helium_2025} propose that the unseen companion is a stripped helium (He) star. Using deep NIRC2 imaging combined with the Keck laser guide star adaptive optics (AO) system, we have performed near-infrared follow-up imaging of PSR J1928+1815 to rule out the hypothesis that the companion is a He star. We summarize our main conclusions below.

\begin{itemize}
    \item We stack NIRC2-LGS exposures acquired over one orbital period of PSR J1928+1815, achieving a 5$\sigma$ detection limit of $K_s \approx 21.3$ at the location of the binary (Figures~\ref{fig:field_comp} and \ref{fig:inject}). Our detection limit is deeper and more robust than previous limits. However, we do not detect any source consistent with the MSP's radio localization.
    \item Combining the spectral models of \citet{gotberg_stripped_2018} with 3D dust maps, we predict the apparent $K_s$-band magnitudes of theoretical stripped stars, ruling out any plausible He star companion (Figure~\ref{fig:theoretical}). We predict apparent $K_s$-band magnitudes of plausible white dwarf (WD) companions, finding that all of them fall below our adopted detection limit (Figures~\ref{fig:theoretical} and \ref{fig:hr_mag}). We consider several hypotheses for the nature of the unseen secondary and rule out scenarios involving either an evolved He star or neutron star companion. We conclude that the companion is very likely a massive WD.
    
    \item Assuming the companion is a massive WD, we consider two possible explanations for the observed radio eclipses: (1) the WD is being ablated, or (2) the WD is young, hot, and is driving its own wind. A recent work by \citet{gong_alternative_2025} favored (1). We find that their assumptions were rather optimistic, and the incident gamma-ray luminosity from the MSP is likely insufficient to ablate the WD and explain the radio eclipses. 
    
    As an alternative, we consider (2). We find that  $\sim$1 GHz radio waves are readily attenuated by synchrotron absorption, and a weak $\dot{M} \sim 10^{-12}$--$10^{-13}\, M_{\odot}$ yr$^{-1}$ wind from a young  WD is sufficient to cause the radio eclipse. Unfortunately, it is observationally and theoretically uncertain whether a young WD can launch such a wind, and if so, for how long. Extrapolating plausible wind mass loss prescriptions, we find that a young, massive WD can drive a sufficiently strong wind for $\sim 10^4$--$10^5$ yr (Figure~\ref{fig:wind}), but this estimate is quite uncertain.
    
    \item We consider all plausible scenarios for the recycling of the NS, and conclude that Case BB mass transfer from a He star companion is the likely formation channel of PSR J1928+1815. \texttt{MESA} simulations of this formation pathway can reproduce the observed properties of the binary, with the He star eventually evolving into a young, hot ONe WD \citep{guo_eclipsing_2025}. In this scenario, the binary emerged from a common envelope as a NS + $\sim 2\,M_{\odot}$ He star binary with an orbital period of $\sim 0.1$\,d. The He star would then have been stripped via stable mass transfer, leaving behind a $\sim 1.2\,M_{\odot}$ WD and recycling the NS.
    
    \item Assuming an (uncertain) formation rate of $2 \times 10^{-4}$ yr$^{-1}$ for eclipsing MSP + young WD binaries in the Milky Way from binary population synthesis \citep[e.g.,][]{guo_eclipsing_2025}, and correcting for beaming, completeness, and selection effects, we find that a $3\times 10^4$ yr detectability lifetime is consistent with detection of $\mathcal{O}(1)$ systems like  PSR J1928+1815. The scenario in which the WD is young and driving its own wind requires observing the system in a short-lived phase of binary evolution, but this may not be a problem, since there are many other MSP + massive WD binaries that are not eclipsing and presumably older \citep{manchester_atnf_2005}.  

\end{itemize}

In the future, ultra-deep near-infrared imaging of the field of PSR J1928+1815 could constrain the presence of the putative WD companion. We predict that a WD young enough to launch an eclipsing wind would have $K_s = 24$--$26$, detectable with SNR $\gtrsim 10$ in a $\gtrsim 1$ ks JWST NIRCam observation \citep{2016jdox.rept......}. Finally, the Chandra X-ray Observatory can be used to search for Doppler-boosted, orbitally modulated synchrotron emission from the intrabinary bow shock \citep{wadiasingh_constraining_2017}.

\begin{acknowledgments}
We thank Hang Gong, Eliot Quataert, Tom Maccarone, Ylva G\"otberg, and Shri Kulkarni for useful discussion. This research was supported by NSF grants AST-2307232 and AST-2508988.  This research was supported in part by grant NSF PHY2309135 to the Kavli Institute for Theoretical Physics (KITP). YG acknowledges support from the National Natural Science Foundation of China (No.\ 12403035). This work has made use of data from the European Space Agency (ESA) mission {\it Gaia} (\url{https://www.cosmos.esa.int/gaia}), processed by the {\it Gaia}
Data Processing and Analysis Consortium (DPAC,
\url{https://www.cosmos.esa.int/web/gaia/dpac/consortium}). Funding for the DPAC
has been provided by national institutions, in particular the institutions
participating in the {\it Gaia} Multilateral Agreement. 
\end{acknowledgments}

\begin{contribution}

PN was responsible for leading the observation, analyzing the data, and writing the manuscript. KE came up with the initial research concept, obtained the funding, and edited the manuscript. JF performed theoretical calculations pertaining to the eclipse mechanism. YG and TMT provided insight on the binary evolution. 


\end{contribution}

%
\facilities{Keck:II (NIRC2-LGS)}

\software{astropy \citep{2013A&A...558A..33A, 2018AJ....156..123A, 2022ApJ...935..167A}}



\appendix

\section{Free-free absorption as an eclipse mechanism}
\label{sec:freefree}

We now show that free-free absorption of the pulsar's radio emission by the WD wind is insufficient to explain the observed eclipse. Following \citet{rybicki_lightman_1986}, the free-free absorption coefficient is given by:

\begin{equation}
    \frac{\alpha_{\text{ff}}}{\text{cm$^{-1}$}} = 0.018  \left(\frac{T}{\text{K}}\right)^{-3/2} \left(\frac{\nu}{\text{Hz}}\right)^{-2}  \left(\frac{Z^2 n_e n_I}{\text{cm$^{-6}$}}\right) \bar{g}_{\text{ff}},
\end{equation}
\vspace{0.1cm}

\noindent where $T$ is the wind temperature, $Z$ is the ion charge, $n_e$ and $n_I$ are the number densities of electrons and ions, $\nu$ is the frequency, and $\bar{g}_{\text{ff}}$ is the velocity-averaged Gaunt factor, respectively. Assuming that the wind consists of fully ionized helium, we have:

\begin{equation}
\begin{aligned}
     \frac{\alpha_{\text{ff}}}{\text{cm$^{-1}$}} &\approx 3.2 \times 10^{21} \left(\frac{T}{10^4\text{ K}}\right)^{-3/2} \left(\frac{\rho}{\text{g cm$^{-3}$}}\right)^2 \\& \left(\frac{\nu}{\text{GHz}}\right)^{-2} \bar{g}_{\text{ff}}.
\end{aligned}
\end{equation}

For simplicity, we take the binary to be edge-on (i.e., $\sin i \approx 1$) and at conjunction, with the WD between the MSP and the observer. Let the gas have typical mass $M$ and length scale $\mathcal{L}$. Suppose that this mass is supplied by a wind with mass loss rate $\dot{M}$ and speed $v_w$, so that $M = \dot{M} \mathcal{L} / v_w$ and $\rho = \dot{M}/{4 \pi \mathcal{L}^2 v_w}$. Then, the optical depth $\tau \sim \alpha_{\text{ff}} \mathcal{L}$ scales as follows:

\begin{equation}
\begin{aligned}
    \tau &\sim 0.025 \left(\frac{\dot{M}}{10^{-12}\,M_{\odot}\text{ yr$^{-1}$}}\right)^2 \left(\frac{v_w}{10^3\text{ km s$^{-1}$}}\right)^{-2} \\ &\left(\frac{\mathcal{L}}{R_{\odot}}\right)^{-3} \left(\frac{T}{10^4\text{ K}}\right)^{-3/2} \left(\frac{\nu}{ \text{GHz}}\right)^{-2} \bar{g}_{\text{ff}}.
\end{aligned}
\end{equation}

Consider a WD of effective temperature $T_{\text{eff, WD}}$, mass $M_{\text{WD}}$ and radius $R_{\text{WD}}$. If the wind is in radiative equilibrium with the WD, then $T(\mathcal{L}) \sim T_{\text{eff, WD}} \left(\mathcal{L}/R_{\text{WD}}\right)^{-1/2}$. Plugging in:

\begin{equation}
\begin{aligned}
    \tau &\sim 0.025 \left(\frac{\dot{M}}{10^{-12}\,M_{\odot}\text{ yr$^{-1}$}}\right)^2 \left(\frac{v_w}{10^3\text{ km s$^{-1}$}}\right)^{-2} \\& \left(\frac{\mathcal{L}}{R_{\odot}}\right)^{-9/4} \left(\frac{T_{\text{eff, WD}}}{10^5\text{ K}}\right)^{-3/2} \left(\frac{R_{\text{WD}}}{0.0055\,R_{\odot}}\right)^{-3/4} \\& \left(\frac{\nu}{1.25 \text{ GHz}}\right)^{-2} \bar{g}_{\text{ff}}.
\end{aligned}
\end{equation}

The length scale $\mathcal{L}$ is set by the standoff between the MSP's wind and the WD's wind, which creates an intrabinary bow shock. The standoff distance from the WD is given by $R_0 = a \sqrt{\eta_w} / (1 + \sqrt{\eta_w})$, where $a$ is the orbital separation and $\eta_w \equiv \dot{M} v_w c / \dot{E}$ is the wind momentum ratio (see Section~\ref{sec:WD_wind}). For the majority of the time that the WD launches a wind, $\eta_w \ll 1$, and we can approximate $\mathcal{L} \sim R_0 \sim a \sqrt{\eta_w}$. Adopting a pulsar spin-down luminosity $\dot{E} = 1.2 \times 10^{35}$ erg s$^{-1}$ and using $a \approx 1.6\,R_{\odot}$, we can write that:

\begin{equation}
\begin{aligned}
    \tau &\sim 13 \left(\frac{\dot{M}}{10^{-12}\,M_{\odot}\text{ yr$^{-1}$}}\right)^{7/8} \left(\frac{v_w}{10^3\text{ km s$^{-1}$}}\right)^{-25/8} \\&\left(\frac{T_{\text{eff, WD}}}{10^5\text{ K}}\right)^{-3/2} \left(\frac{R_{\text{WD}}}{0.0055\,R_{\odot}}\right)^{-3/4} \left(\frac{\nu}{1.25 \text{ GHz}}\right)^{-2} \bar{g}_{\text{ff}}.
\end{aligned}
\end{equation}

\vspace{0.2cm}
Finally, the wind speed is set by the escape velocity of the WD, such that $v_w \sim \sqrt{2 G M_{\text{WD}}/R_{\text{WD}}}$. Plugging in, we find that:

\begin{equation}
\begin{aligned}
    \tau &\sim 0.01 \left(\frac{\dot{M}}{10^{-12}\,M_{\odot}\text{ yr$^{-1}$}}\right)^{7/8} \left(\frac{M_{\text{WD}}}{1.2\,M_{\odot}}\right)^{-25/16} \\ &\left(\frac{T_{\text{eff, WD}}}{10^5\text{ K}}\right)^{-3/2} \left(\frac{R_{\text{WD}}}{0.0055\,R_{\odot}}\right)^{13/16} \left(\frac{\nu}{1.25 \text{ GHz}}\right)^{-2} \bar{g}_{\text{ff}}.
\end{aligned}
\end{equation}

\vspace{0.1cm}
In other words, for typical WD parameters, we find that $\tau \ll 1$ at $L$-band radio frequencies. We conclude that free-free absorption of the pulsar's radio emission by a young, hot (proto-)WD's $\dot{M} \gtrsim 10^{-12}\,M_{\odot}$ yr$^{-1}$ wind is insufficient to explain the observed radio eclipse.


\bibliography{bibliography}{}
\bibliographystyle{aasjournalv7_modified}



\end{document}